\documentclass[twocolumn,apl]{revtex4}
\usepackage{amsfonts}
\usepackage[T1]{fontenc}
\usepackage{amsmath,amsbsy,amssymb,graphicx}
\usepackage{times}
\usepackage{color}
\let\mathbf=\boldsymbol

\begin{document}

\title{{\Large Purely electrical detection of a skyrmion\\}
{\Large in constricted geometry}}
\author{Keita Hamamoto$^1$, Motohiko Ezawa$^1$ and Naoto Nagaosa$^{1,2}$}
\affiliation{$^1$Department of Applied Physics, University of Tokyo, 7-3-1 Hongo, Bunkyo-ku, Tokyo 113-8656, Japan}
\affiliation{$^2$RIKEN Center for Emergent Matter Science (CEMS), Wako, Saitama 351-0198, Japan}

\begin{abstract}
How to detect the skyrmion position is a crucial problem in future skyrmionics since it corresponds to the reading process of information. We propose a method to detect the skyrmion position purely electrically by measuring the Hall conductance in a constricted geometry. The Hall conductance becomes maximum when a skyrmion is at the lead position. It is possible to detect the skyrmion position even at room temperature. We find an optimized width of the sample determined by the skyrmion radius. We also investigate the effects of elastic and inelastic scatterings, and finite temperature. We find that the local density of states become minimum at the skyrmion position. Our results will be a basis of future skyrmion electronics.
\end{abstract}

\maketitle

{\it Introduction}: A magnetic skyrmion, a topological excitation of spin texture, is experimentally observed in many itinerant ferromagnets without inversion symmetry such as B20 compounds. MnGe, MnSi, FeGe are typical examples of this category. This particle-like object is expected to be the basis of future skyrmionics, where skyrmions are used for the memory and information processes\cite{SkRev,Skmemory}. In practical application of a skyrmion, writing and reading processes are inevitably important. There are many proposals on the writing process of a skyrmion. A skyrmion is created by applying a spin polarized current\cite{Sampaio}, a circulating spin current\cite{Han}, laser beam\cite{Marco,GSk,Koshibae}, from a notch\cite{IwasakiN} and converted form a domain wall pair\cite{Yan}. Compared with these creation methods, the detection of a single skyrmion position is rather difficult. One method is using the Lorentz TEM (Tunneling Electron Microscopy)\cite{Yu}. However, the experimental apparatus is very expensive, and it is hard to make an observation of short-time dynamics. Recently another method is proposed experimentally\cite{Hanneken} and theoretically\cite{Crum}, where the tunneling magnetoresistance is used. A purely electric detection of the skyrmion position is desirable for future nanoelectronics applications of skyrmions.

A prominent feature of a skyrmion is that it produces emergent magnetic field, which originates from the solid angle subtended by the spins called scalar spin chirality. The topological property of skyrmion guarantees that the total flux generated by one skyrmion is one flux quantum, $h/e$. The size of a skyrmion is $\sim$1nm for atomic Fe layer on Ir(111) surface~\cite{Heinze}, $\sim$3nm for MnGe~\cite{Kanazawa}, $\sim$18nm for MnSi~\cite{Lee}, and $\sim$70nm for FeGe~\cite{FeGe}. The corresponding emergent magnetic field is $\sim 4000$T, $1100$T, $28$T, and $1$T, respectively. The topological Hall effect is a manifestation of emergent magnetic field, where the Hall effect occurs in the presence of skyrmions\cite{Lee,Schulz,Li,Kanazawa,Neu}. Moreover, the quantized topological Hall effect has been theoretically proposed\cite{Hamamoto}.

In this paper, we propose a method of detecting the skyrmion position in conducting systems such as B20 compounds purely by an electric method. We investigate the Hall conductance based on the Landauer B\"{u}ttiker formula by attaching the side leads. The Hall conductance is found to have a peak where a skyrmion is located at the lead position, while it reduces when a skyrmion is away from the lead. By detecting the peak of the Hall conductance, it is possible to determine when a skyrmion passes through the vicinity of the lead even at room temperature. We investigate various sample and lead widths and find an optimized sample geometry for a skyrmion radius. We also study the effect of elastic and inelastic scatterings, and finite temperature. Furthermore we show that we can determine the position of skyrmion by measuring the local density of states using the STM (Scanning Tunneling Microscopy).

\begin{figure}[!t]
\centerline{\includegraphics[width=0.5\textwidth]{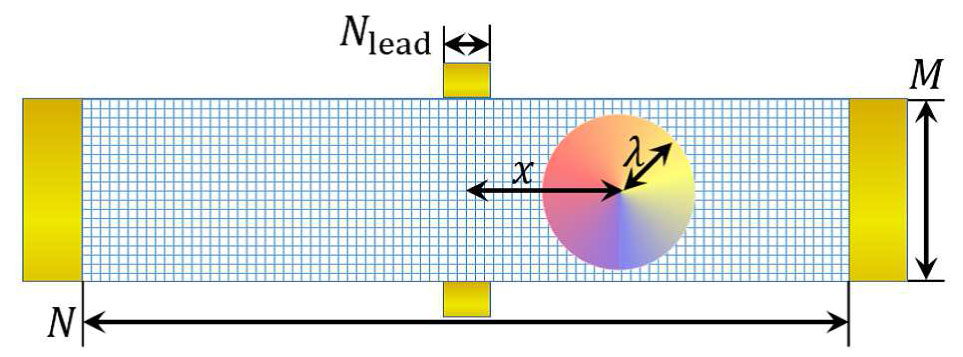}}
\caption{(a) Illustration of the system which consists of the sample and four leads. The sample size is the length $N$ times the width $M$. We attach four leads, where two leads are attached on left and right sides and the other two Hall leads are attached at the central region. The width of the Hall leads is $N_{\text{lead}}$. A skyrmion exists in the sample, in which the center position of the skyrmion is set to be $x$. The skyrmion radius is $\lambda$.}
\label{FigIllust}
\end{figure}

{\it Model}: Figure 1 illustrates the system we investigate in the present paper. We consider a sample made of a nanoribbon of the square lattice. Noninteracting electrons are coupled with the background spin texture\cite{Anderson,Ohgushi,Onoda,Hamamoto}. Originally, the system is described by the double-exchange model. By taking the strong coupling limit, we obtain the effective tight-binding Hamiltonian as\cite{Anderson,Ohgushi,Onoda,Hamamoto}
\begin{equation}
H_{\text{D}}=-\sum_{ij}t^{ij}_{\text{eff}}d^{\dagger}_id_j ,
\label{Hamil}
\end{equation}
where $d^{\dagger}_i$ ($d_i$) is the creation (annihilation) operator of a electron at the $i$ site, whose spin is forced to align with the local spin texture. The effective transfer integral is given by
\begin{equation}\label{teff}
t^{ij}_{\text{eff}} =te^{ia_{ij}}\cos\frac{\theta_{ij}}{2},
\end{equation}
where $a_{ij}$ is the emergent gauge field induced by the skyrmion and $\theta_{ij}$ is the angle between two spins.
We assume the skyrmion profile as $\theta (\mathbf{r})=\pi (1-r/\lambda )$ 
for $r<\lambda$ and $\theta (\mathbf{r})=0$ for $r>\lambda$
in numerical analysis in what follows.

\begin{figure}[!t]
\centerline{\includegraphics[width=0.5\textwidth]{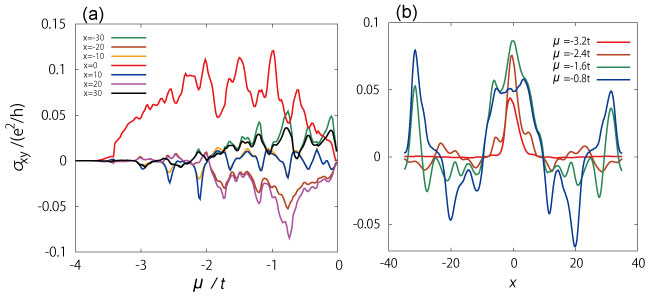}}
\caption{(a) Hall conductance as a function of the chemical potential for various skyrmion positions. The horizontal axis is the chemical potential $\mu$, while the vertical axis is the Hall conductance. (b) Hall conductance as a function of the skyrmion position $x$ for different chemical potentials.}
\label{FigConducEF}
\end{figure}

{\it Landauer-B\"{u}ttiker formula}: The Hall conductance is calculated based on the following Landauer-B\"{u}ttiker formalisms.\cite{Datta} The current flowing into the $p$-th lead is expressed as $I_{\text{p}}=\sum_{q=1}^4 G_{\text{p,q}}(V_{\text{p}}-V_{\text{q}})$, where 
\begin{equation}
G_{\text{p,q}} (E_{\text{F}})=(e^{2}/h) T_{\text{p,q}} (E_{\text{F}})
\end{equation}
for zero temperature and
\begin{equation}
G_{\text{p,q}} (E_{\text{F}})=(e^{2}/h) \int \mathrm{d}E \ T_{\text{p,q}} (E) \left(-\frac{\partial f}{\partial E}\right)
\end{equation}
for finite temperature with $f$ being the Fermi distribution function. The transmission probability is
\begin{equation}
T_{\text{p,q}} (E)=\text{Tr}[\Gamma _{\text{p}}(E)G_{\text{D}}^{\dag}(E)\Gamma _{\text{q}}(E)G_{\text{D}}(E)], \label{G}
\end{equation}
where $\Gamma _{\text{p}}(E)=i[\Sigma _{\text{p}}(E)-\Sigma _{\text{p}}^{\dag }(E)]$ with the retarded self-energies due to the $p$-th lead, $\Sigma _{\text{p}}(E)$, and
\begin{equation}
G_{\text{D}}(E)=\left[ E-H_{\text{D}}-\Sigma _{\text{inela}} -\sum_{p}\Sigma _{\text{p}}(E) \right] ^{-1},  \label{StepA}
\end{equation}
with the Hamiltonian $H_{\text{D}}$ in Eq.(\ref{Hamil}) for the device region.
We attach square-lattice semi-infinite leads as shown in Fig.\ref{FigIllust}.
The analytical expression of the self-energy of the square-lattice lead is known\cite{Datta}.
The Hall resistance $R_\text{H}$ and the longitudinal resistance $R_\text{L}$ can be calculated from $G_{\text{p,q}} (E)$ as
\begin{equation}
R_\text{H}=(V_4-V_3)/I_1 , \quad R_\text{L}=(V_1-V_2)/I_1
\end{equation}
under the conditions, $I_2=-I_1 , I_3=I_4=0$. The Hall conductance is $\sigma_\text{xy} =-R_\text{H}/(R^2_\text{H}+R^2_\text{L})$ and the longitudinal conductance is $\sigma_\text{xx} =R_\text{L}/(R^2_\text{H}+R^2_\text{L})$.
The self-energy $\Sigma _{\text{inela}}=-i\hbar/2\tau \equiv -i\eta $ is considered to handle the effects of inelastic scatterings such as electron-electron and electron-phonon interactions, which are inevitable in realistic situations. 

In the following, we set $N=81, M=12, N_{\text{lead}}=3, \lambda=5, \eta=0.04t, E_{\text{F}}=-3.2t$ unless otherwise specified, which are approximately the optimized values as discussed below.

\begin{figure}[!t]
\centerline{\includegraphics[width=0.3\textwidth]{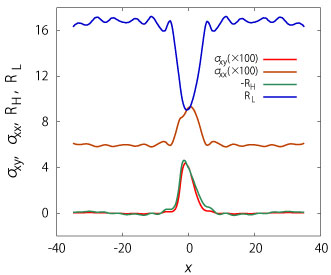}}
\caption{Dimensionless Hall conductance, Hall resistance, and longitudinal conductance and longitudinal resistance as a function of the skyrmion position.}
\label{Figres}
\end{figure}

{\it Hall conductance}: We show the Hall conductance as a function of the chemical potential by changing the skyrmion position in Fig.\ref{FigConducEF}(a).
The Hall conductance shows a complicated structure as a function of the chemical potential.
However, there is a simple structure in the vicinity of the band edge, and the Hall conductance is finite only around $x=0$ in this region. In Fig.\ref{FigConducEF}(b), we show the Hall conductance as a function of the skyrmion position $x$. We find the significant peak structure at $x\sim 0$ when $E_{\text{F}}=-3.2t$, which suggests that one can detect the position of a skyrmion by measuring the Hall conductance in metallic materials with low carrier concentration. In the band edge, the band structure of the square lattice is well described by the free electron band. We concentrate on this region. In Fig.\ref{Figres}, we show the Hall resistance $R_\text{H}$, the longitudinal conductance $\sigma_\text{xx}$ and the longitudinal resistance $R_\text{L}$ which show similar peak structures when the skyrmion comes close to the Hall leads. Especially, the Hall resistance, being often probed in usual Hall measurement, shows almost the same behaver as the Hall conductivity. Hereafter, we only show the Hall conductivity.

The origin of the peak structure is naturally understood since the electron wave function is scattered by the emergent magnetic field induced by the skyrmion. In a semi-classical picture, an electron wave packet feels the Lorentz force. If the scattered wave packet enters the side lead, there is a Hall conductance. The wave packet strongly turns at the skyrmion since the emergent magnetic field is very strong.

\begin{figure}[!t]
\centerline{\includegraphics[width=0.5\textwidth]{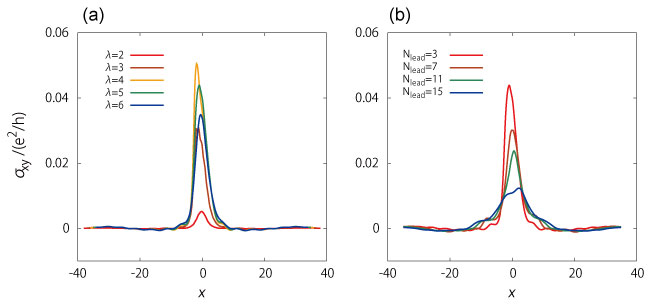}}
\caption{(a) Hall conductance as a function of the skyrmion position for various skyrmion radius. (b) Hall conductance for $\lambda=5$ as a function of the skyrmion position for different width of Hall leads, $N_{\text{lead}}$.}
\label{FigrskNlead}
\end{figure}

We show the Hall conductance for various skyrmion radius by fixing other parameters in Fig.\ref{FigrskNlead}(a).
The peak value of the Hall conductance depends on the skyrmion radius, which implies there is an appropriate skyrmion size to detect the signal. This is understood as follows. For a small skyrmion, almost all electrons do not feel the emergent magnetic field since the region of the magnetic field is small although the magnetic field is large. On the other hand, if a skyrmion is large, the emergent magnetic field is weak although almost all electrons feel the magnetic field. As a result, the Hall conductance takes the maximum value for a finite skyrmion size. The Hall conductance for various values of Hall lead width $N_{\text{lead}}$ is shown in Fig.\ref{FigrskNlead}(b). The peak becomes small and broad as Hall leads become wide, which implies that narrower Hall leads are better for accurate detection of the position of a skyrmion.

\begin{figure}[!t]
\centerline{\includegraphics[width=0.5\textwidth]{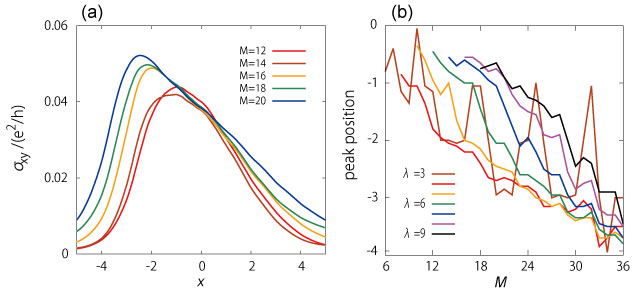}}
\caption{ (a) Hall conductance for $\lambda =5$ as a function of the skyrmion position for various sample width $M$ in the vicinity of $x=0$. (b) Position of the peak as a function of sample width $M$ for various values of skyrmion radius $\lambda$.}
\label{FigM}
\end{figure}

In realistic situation, the skyrmion radius is usually fixed for each material and is hard to control experimentally, and therefore the optimization of experimentally tunable parameters for a given value of skyrmion radius is required. For example, the device width $M$ is rather easy to control. In Fig.\ref{FigM}(a), we show the Hall conductance in the vicinity of the peaks for various device width $M$. We find a systematical shift of the peak position as a function of $M$. Figure \ref{FigM}(b) shows the peak position as a function of $M$. It is clearly seen that the peaks shift almost linearly as $M$ increases. At the same time, the peak height becomes large as $M$ increases. These trends are not the case for $\lambda =3$ where the discreteness of skyrmion spin configuration is relevant. These results indicate that a wide device will enlarge the signal peak, but the accuracy of the detected skyrmion position will be lowered. 

The linear shift of the peak can be understood as follows. The electron wave packet scattered by the skyrmion has a finite ``scattering angle'' $\phi$, and the lateral distance between the scattering center and the Hall lead is $M/2$. Thus, the longitudinal distance, i.e., the peak shift is approximated as $M/2\tan{\phi} \propto M$. The ``scattering angle'' evaluated from the slope of Fig.\ref{FigM}(b) for $\lambda=4$ is about $\phi \sim 80^\circ$ and has small dependence on the radius. This value is larger than $60^\circ$, which is the maximum value of the scattering angle of magnon discussed in Refs.\cite{Iwasaki,Schutte}. This discrepancy may attribute to two reasons. One is the complicated interference of the wave function due to the reflection at the edge of the device. The other is that the incident wave of electrons has a finite $y$-direction momentum $k_y$ because the injected electrons occupy a state within the Fermi circle. This nonzero $k_y$ results in the enhancement of the ``scattering angle'' although the mean value of $k_y$ is zero. On the other hand, only the $k_y=0$ component of the incident wave is considered in Ref.\cite{Iwasaki}. The fact that the shift of the peak is tiny as a consequence of large and constant ``scattering angle'' is advantageous for the accurate determination of the skyrmion position.

\begin{figure}[!t]
\centerline{\includegraphics[width=0.5\textwidth]{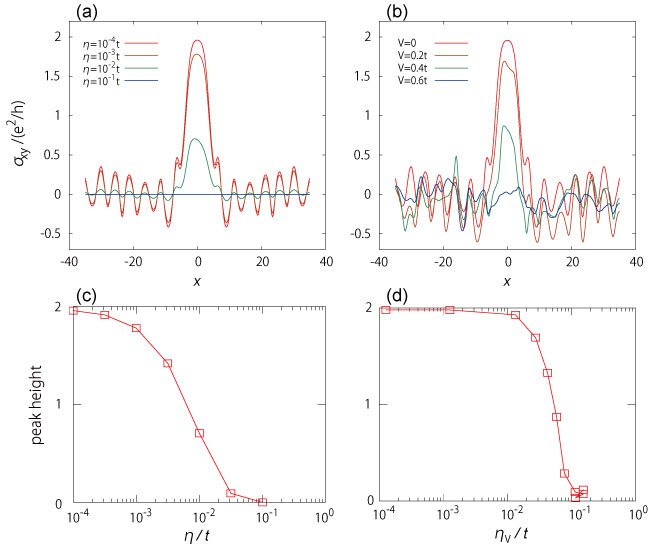}}
\caption{Hall conductance as a function of the position (a) for various inelastic scattering constants $\eta$, (b) for various elastic disorder strengths $V$. Here, we set $\eta=10^{-4}t$. Peak height as a function of (c) the inelastic scattering constant $\eta$, (d) the elastic scattering constant $\eta_{\text V}$ (see the main text for the definition.).}
\label{Figdisorder}
\end{figure}

\begin{figure}[!t]
\centerline{\includegraphics[width=0.5\textwidth]{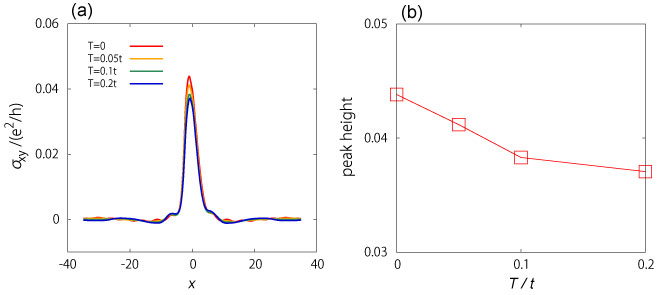}}
\caption{(a) Hall conductance at finite temperature. (b) Temperature dependence of the peak height.}
\label{FigfiniteT}
\end{figure}

{\it Effects of disorders}: We show the Hall conductance for various inelastic scattering strength $\eta$ in Fig.\ref{Figdisorder}(a). 
The peak height around $x=0$ reduces as $\eta$ increases.
There are oscillating tails of the Hall conductance as a function of the skyrmion position away from the peak, which are smeared by the strong inelastic scattering strength.

In order to investigate the effect of local impurity and disorder, we introduce the disorder potential as $H_{\text{imp}}=\sum_{i}U_{i}d^{\dagger}_id_i $, where $U_i$ is a uniform random number ranging $-V<U_i<V$. We have calculated the Hall conductance for various disorder strength $V$ shown in Fig.\ref{Figdisorder}(b). We can clearly see that the peak height become smaller as the disorder become stronger. However the peak does not disappear even with the strong disorder $V\simeq 0.4t$.

To compare the effect of the inelastic and the elastic disorders, we introduce the elastic scattering constant $\eta _{\text V}$ as $\eta _{\text V}\equiv \hbar / 2\tau _{\text V} \equiv \pi n_\text{imp}\left< |U_i|^2 \right> \rho (E_F)$ with $n_\text{imp}=1$ being the density of scatterers, $\rho (E_F)$ is the density of states at the Fermi energy and $\left< |U_i|^2 \right>=V^2/3$ is the expectation value over the uniform stochastic distribution of the disorder potential. In Figs.\ref{Figdisorder}(c) and (d), we show the inelastic and the elastic scattering constant dependence of the peak height, respectively. It can be seen that the peak structure is less robust against inelastic scatterings than elastic ones.

Furthermore, the Hall conductance at finite temperature is shown in Fig.\ref{FigfiniteT}. The signal peak never disappear even when the temperature is an appreciable fraction ($\sim0.2$) of the transfer integral $t\sim 1\text{eV}\sim 10^4\text{K}$. This extreme robustness against thermal effect may be discussed as follows. In the zero temperature, the energy dependence of the Hall conductance is not so strong for $\mu\sim -3.2t$ and even the energy-integrated value is quite large for $x=0$ compared to other positions as shown in Fig.\ref{FigConducEF}. Thus the energy integration in the finite temperature has only a small effect on the peak structure.

These results suggest that the peak structure is quite robust against the presence of various kinds of disorders such as electron-electron scatterings, electron-phonon scatterings, elastic impurities and thermal smearing. Thus, our method of skyrmion detection does not require extremely high quality sample fabrication nor cooling processes. 

\begin{figure}[!t]
\centerline{\includegraphics[width=0.5\textwidth]{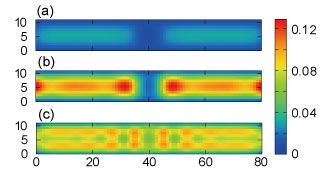}}
\caption{Real space mapping of the local DOS in the presence of a skyrmion at the center for (a) $E_F=-4.0t$, (b) $E_F=-3.9t$, (c) $E_F=-3.8t$.}
\label{FigDOS}
\end{figure}

{\it Local density of states}: The local density of states (DOS) at the $i$ site is given by
\begin{equation}
\rho _{i}(E)=-\pi ^{-1}\text{Im}[G_{\text{D}}(E)_{ii}],
\end{equation}%
in terms of the Green function $G_{\text{D}}(E)$ of the device.

We show the local DOS of the nanoribbon where a skyrmion exists at the center in Fig.\ref{FigDOS}.
The local DOS substantially reduces around the skyrmion when the chemical potential is in the vicinity of the band bottom. We have confirmed that the reduction of local DOS is due to both the phase factor and the cosine factor in the effective transfer integral (See Eq.\ref{teff}) and we have checked the effect of the cosine factor is about three times larger. The effective band width is reduced around the skyrmion hence the low energy electrons cannot enter this region when $-4t\leq E_F \lesssim -4|t_\text{eff}|$ with $t_\text{eff}$ being the typical value of the effective transfer integral inside the skyrmion region. This result suggests that we can determine the skyrmion position by observing the local DOS by the STM \cite{Hanneken}.

{\it Discussion}: We have proposed a method to detect a skyrmion by measuring the Hall conductance in conducting material. The peak of Hall conductance appears when the chemical potential lies near the band bottom. Hence, our detection method favors the material with low carrier concentration. The merit of this method is that it needs less costs compared with other methods such as the Lorentz TEM. Another merit is that the real-time observation of the skyrmion position is possible since we can observe very rapid change in the Hall conductance. We have considered the static skyrmion in the present paper, but the time-scale of the electron is much faster than that of the skyrmion motion. The former is typically $\sim \frac{\hbar}{0.1t} \sim 0.1 \text{ps}$ while the latter is $\sim 0.1-1 \text{ns}$. Therefore, this approximation is justified. The accuracy of the detected position is quite high due to the anomalously large ``scattering angle'', whose microscopic origin is yet to be explored. Our detection method of skyrmion will be a promising candidate mechanism of reading process in skyrmion based memory devices such as a skyrmion racetrack memory where a skyrmion is driven by electric current on a circuit.\cite{Skmemory} Alternatively, we found that the skyrmion position can be determined by measuring the local DOS, which will be possible by the STM. Our results will be a clue for future nanoelectronics devices composed of skyrmions.

This work was supported in
part by JSPS KAKENHI Grants No. 24224009, No. 25400317, No. 15H05854, and 
No. 26103006.

\end{document}